# Synthesizing Products for Online Catalogs

Hoa Nguyen*
University of Utah
thanhhoa@cs.utah.edu

Ariel Fuxman
Microsoft Research
arielf@microsoft.com

Stelios Paparizos
Microsoft Research
steliosp@microsoft.com

Juliana Freire
University of Utah
juliana@cs.utah.edu

Rakesh Agrawal
Microsoft Research
rakesha@microsoft.com

## ABSTRACT

A comprehensive product catalog is essential to the success of Product Search engines and shopping sites such as Yahoo! Shopping, Google Product Search, and Bing Shopping. Given the large number of products and the speed at which they are released to the market, keeping catalogs up-to-date becomes a challenging task, calling for the need of automated techniques. In this paper, we introduce the problem of product synthesis, a key component of catalog creation and maintenance. Given a set of offers advertised by merchants, the goal is to identify new products and add them to the catalog, together with their (structured) attributes. A fundamental challenge in product synthesis is the scale of the problem. A Product Search engine receives data from thousands of merchants about millions of products; the product taxonomy contains thousands of categories, where each category has a different schema; and merchants use representations for products that are different from the ones used in the catalog of the Product Search engine.

We propose a system that provides an end-to-end solution to the product synthesis problem, and addresses issues involved in data extraction from offers, schema reconciliation, and data fusion. For the schema reconciliation component, we developed a novel and scalable technique for schema matching which leverages knowledge about previously-known instance-level associations between offers and products; and it is trained using automatically created training sets (no manually-labeled data is needed). We present an experimental evaluation using data from Bing Shopping for more than 800K offers, a thousand merchants, and 400 categories. The evaluation confirms that our approach is able to automatically generate a large number of accurate product specifications. Furthermore, the evaluation shows that our schema reconciliation component outperforms state-of-the-art schema matching techniques in terms of precision and recall.

## 1. INTRODUCTION

With the emergence of online shopping there has been a surge of commercial portals and comparison shopping sites, such as Amazon, Ciao, and PriceGrabber. Recently, search engine companies have entered this market with Product Search Engines or shopping verticals such as Yahoo! Shopping, Google Product Search, and Bing Shopping. The success of these systems heavily relies on the variety and quality of the products that they present to users. In that sense, the product catalog is to online shopping what the Web index is to Web search. The catalog of a Product Search Engine contains products represented in a structured format. For example, a digital camera may have attribute-value pairs describing its resolution, focal length, etc. This structured data is fundamental to drive the user experience: it enables faceted search, comparison of products based on their specifications, and ranking of products based on their attributes. While for Web search engines it is well understood how to automatically gather and index unstructured Web pages, Product Search Engines face new challenges in that they need to *extract and integrate structured data at a Web scale*. Existing approaches to catalog construction require substantial manual effort, leading to critical limitations in terms of product coverage and freshness.

A Product Search Engine is an intermediary between online buyers and merchants who sell the products. Figure 1 shows a screen shot of a page from a Product Search Engine describing a product and associated offers from merchants who sell the product. Upon clicking on one of the links in an offer, the user is redirected to the Web page of the merchant, where she can purchase the product. Merchants provide offers to the Product Search Engine, which can then be connected to products in the catalog using universal identifiers (if available) or an offer-to-product matching algorithm.

Clearly, if a product does not exist in the catalog, the merchant offers for the product will never be shown. This results in loss of revenue for the merchant, who is missing a possible transaction; and for the Product Search Engine, which is paid based on clicks. In this paper, we invert the equation and propose the following: If a product is missing from the catalog, rather than dropping the offers, use them to construct a product representation that can be added to the catalog. This problem is challenging for multiple reasons. Merchants provide offer feeds which often contain minimal structured data: a title (short unstructured sentence such as "HP 400GB 10K 3.5 DP NSAS HDD"); a price; and the URL of the landing page on the merchant site where the product can be bought (see Figure 3). We overcome this problem by leveraging the landing page of the offer. In fact, even when the feeds do not contain enough information about the specifications, the landing page usually contains the product specifications in a structured format, *i.e.,* containing the set of attribute-value pairs that describe the product (see for example, the two landing pages in Figure 1). Since extraction of product information from merchant Web pages must be done in a fully automated fashion, without making any category- or merchant-specific assumptions, invariably, this will lead to noisy data. Thus, product

*Work done as an intern at Microsoft

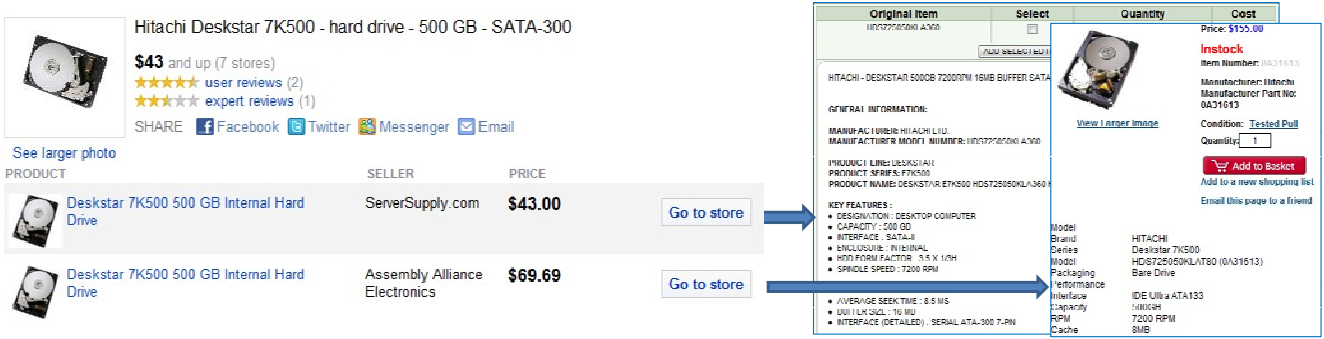

Figure 1: The product snippet returned by a Product Search Engine (left) points to a set of offers from different merchants (right).

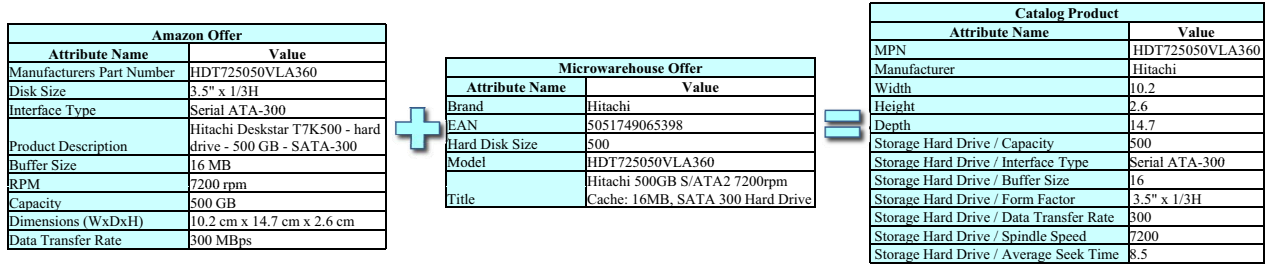

Figure 2: Fusing offers to create product instances.

synthesis must be carried out in such a way that it can handle noise in the information extracted from offers.

Regardless of whether the attribute-value pairs for a product come from a merchant offer in feeds or landing pages, another major problem we must address is the heterogeneity of offers across different merchants. Consider, for example, the hard drive offers in Figure 2 obtained from two merchants: Amazon and Microwarehouse. The Amazon offer contains nine attributes, while the Microwarehouse offer contains only five. Different names are used to represent the same attribute in the two offers (*e.g., Hard Disk Size* and *Capacity*). Similarly, different representations are used for the same value (*e.g., 500* and *500GB*). This problem is compounded by the fact that even within a product category (*e.g.,* hard drive, laptop), there can be a large number of merchants, each using a different schema.

**Contributions.** We present an end-to-end solution to the product synthesis problem. Our main contributions include:

- A system that implements the product synthesis pipeline. Given a set of merchant offers, the system *automatically* derives a set of product instances in a structured format that is compatible with the schema of products in the catalog.
- A novel, scalable method for schema reconciliation which caters to the specific requirements of product synthesis. The method leverages historical knowledge about offer-product associations as well as the large number of offers and different merchants to improve the quality of matches based on the distribution of attribute values. We employ a learning classifier to identify new associations, and we propose a strategy to construct the training sets for the classifier automatically. This makes our approach scalable and able to handle a (very) large number of categories in a catalog taxonomy.
- A thorough experimental evaluation of the end-to-end product synthesis system using data from Bing Shopping for more than 800,000 offers, 1,000 merchants, and 400 categories. The evaluation shows that our approach is able to generate a large number of accurate product instances. We also show that schema matching approach compares favorably against state-of-the art techniques.

## 2. PRODUCT SYNTHESIS: OVERVIEW

A Product Search Engine (PSE) has a *catalog* that consists of a set of *product instances*, and a *product taxonomy* with categories such as laptops and digital cameras. Each product is associated to exactly one category in the taxonomy, which in turn, is represented by a schema that contains a set of attributes. For example, the schema for category "Digital Cameras" contains attributes "Resolution", "Size", etc. Each product $p$ in category $C$ is represented as $p = (C, \{\langle A_1, v_1 \rangle, \ldots, \langle A_n, v_n \rangle\})$, where the attribute-value pairs $\langle A_i, v_i \rangle$ correspond to the product specification and each attribute name $A_i$ belongs to the schema of category $C$.

*Merchants* provide *offers* to the PSE through feeds. An offer $o$ for a product $p$ sold by a merchant $M$ is represented by a tuple $o = (M, price, image, C, URL, title, \{\langle A_1, v_1\rangle, \ldots, \langle A_n, v_n\rangle\})$. It contains information about price, an image, and the URL of the landing page where the product can be bought. It also includes a title which consists of a short free-text sentence describing the product. Furthermore, each offer has a set of attribute-value pairs $\langle A_i, v_i \rangle$, that we refer to as *offer specification*, which describe the product being sold (although they may not conform to the catalog schema for category $C$). Also, although the merchants do not necessarily expose a schema, in the sequel we will abuse the terminology and say that $A_1, \ldots, A_n$ are attributes in the "schema" of merchant $M$ for category $C$.

Merchant feeds may not have category information, or they may have categories under a taxonomy that is different from the one used in the catalog. To determine the category for a given offer, we use a simple classifier, which given the title of the offer, returns its category $C$ under the catalog taxonomy. Due to space limitations, we omit a detailed description of the classifier. We should note, however, that the product synthesis pipeline is resilient to errors in the classifier as long as there is a sufficient number of (representative) offers associated with the product to be synthesized.

Given a set of offers, the goal of *product synthesis* is to create product instances that are compatible with the schemata of the categories in the catalog. Figure 4 shows the high-level architecture of the end-to-end framework that we propose to achieve this goal. The process consists of two main phases: Offline Learning and Run-Time Offer Processing. At the heart of the *Offline Learning* phase is the *Attribute Correspondences Creation* component, which produces correspondences between attribute names in offer specifications and attribute names in the catalog schemata. For example, in Figure 2, the *Hard Disk Size* attribute in the Microwarehouse offer corresponds to the attribute *Capacity* in the catalog schema for hard drives. We construct a classifier that learns to identify correspondences from historical offers matched to products in the catalog (Section 3). While the offer specification may be provided in the feed, in practice, most feeds contain little structured data (see Figure 3). To obtain the offer specification, we exploit the URL to the merchant page which contains the offer. This information is usually presented in a structured fashion (*e.g.,* formatted as a table) and can be automatically extracted. This extraction is done by the *Web-page Attribute Extraction* component, which produces a set of attribute-value pairs for the offer.

| Source Url | Title | Description | Price | Seller | Category |
|---|---|---|---|---|---|
| http://www.tech forless.com/… | Gear Head DVD+/- … | Supports direct -to-disc … | $67 | Tech for Less | Computing\|Stor age\|Hard Drives |
| http://www.lacc .com/... | HP HDD ... | Ref. 3.2gb Ide 3.5 hdd … | $71 | lacc.com | Computing\|Stor age\|Hard Drives |
| http://www.lacc .com/… | TM1280S Hard Quantum Drives… |  | $128 | lacc.com | Computing\|Stor age\|Hard Drives |

**Figure 3: Fragment of an offer feed.**

The *Run-Time Offer Processing Pipeline* creates new products using offers that cannot be matched to any existing product in the catalog (Section 4). In the pipeline, the *Web Page Attribute Extraction* component (also used in the offline phase) produces attribute-value pairs for the offers. The *Schema Reconciliation* component applies the attribute correspondences learned in the Offline Learning phase to translate attribute names of the offers into names that are present in the catalog schema. The reconciled offers are then grouped by the *Clustering* component, where each cluster corresponds to exactly one product. Finally, the product clusters go through the *Value Fusion* component which creates a single product specification for each cluster.

## 3. OFFLINE LEARNING

The goal of the Offline Learning phase is to create a classifier that determines attribute correspondences. We define the notion of attribute correspondence as follows.

DEFINITION 1. *Let $C$ be a category, and $M$ a merchant. Let $A^p$ be an attribute from the catalog schema for category $C$. Let $A^o$ be an attribute from the schema of merchant $M$ for category $C$ (i.e., $A^o$ appears in offers of $M$ for category $C$). We say that $\langle A^p, A^o, M, C \rangle$ is an attribute correspondence from $A^p$ to $A^o$ for category $C$ if $A^p$ and $A^o$ have the same meaning in category $C$.*

We leverage known associations between offers and products to compute the similarity between attribute values. Rather than using all attribute values, similarity is computed only for values that appear in an existing association (Section 3.1). The derived similarities are used as the classifier feature set. As we describe in Section 3.2, the training set for the classifier is constructed in a fully automated fashion, based on practical assumptions for the product synthesis problem.

### 3.1 Using Historical Offer-to-Product Matches

A distinctive aspect of our approach is that it leverages historical knowledge about offer and product associations to identify new attribute correspondences. The business model of Product Search Engines implies the existence of a wealth of historical information about merchant offers associated ("matched") to catalog products. As Figure 1 illustrates, a Product Search Engine exposes merchant offers that are *associated* with a given product; and revenue is derived from user clicks on these offers. Associations between offers and products can be obtained through various methods, including the use of universal identifiers (GTIN, UPC, EAN) when available, manual techniques, or automated matchers that attempt to match the title of the offers to structured product records. We refer to these associations as *historical offer-to-product matches*.

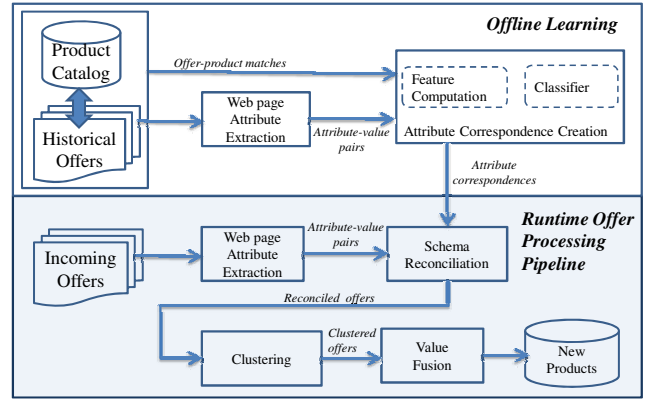

**Figure 4: Product Synthesis Architecture.**

Various schema matching methods in the literature rely on the observation that related attributes will contain similar values (*e.g.,* [10]). However, little attention has been given to the use of historical instance-based matches to compute similarity measures. Here we leverage this information to identify the correspondences between attributes from offers and catalog which have *similar value distributions*. Consider the scenario in Figure 5(a). On the left, we have a list of hard drives from the catalog, and on the right, a list of hard drive offers from one particular merchant. Notice that attribute names alone are not sufficient to derive matches, as the vocabulary used in the merchant and the product catalog schemata are quite different. For example, the speed of the hard drive is referred to as "Speed" in the catalog, and "RPM" in the merchant offer. Although value distributions can be helpful here, their direct use (as done in previous works) may be problematic in the context of product synthesis. The reason is that the type of products provided by a merchant (and thus, the value distributions) may be quite different from those in the catalog. For example, the merchant "SonyStyle.com" only provides Sony MP3 players, but the product catalog may have many different brands of MP3 players, and thus, different distributions for the Brand attribute of the MP3 player category. As another example, in the scenario of Figure 5(a), it might be difficult to detect that Speed and RPM are synonyms because there are some products in the catalog with a speed of 10,000 rpm, and none in the merchant offers.

To address this problem, our Attribute Correspondence Creation component obtains value distributions only from offers and products that match to each other. For offers associated to a given product, as our experimental evaluation shows, it is reasonable to assume that similar attributes will lead to similar value distributions. Consider Figure 5(b): after we remove the products that do not match any offer, the distribution of values for Speed and RPM turn out to be exactly the same, and this enables us to leverage value distributions for inferring accurate attribute correspondences.

**Computing distributional similarity features.** We use a bag of words to collect the values of each attribute in catalog products as well as for merchant offer specifications. To compare distributions, we draw upon work on *distributional similarity* that originates from the Natural Language Processing community [12, 18], where distributional similarity has been used to find synonymous

## Figure 5: Deriving attribute correspondences

**(a) Matching product instances and offers**

| Brand | Model | Speed | Interface |
|---|---|---|---|
| Seagate | Barracuda | 5400 | ATA 100 |
| Seagate | Cheetah | 10000 | ATA 100 |
| Western Digital | Raptor | 7200 | IDE 133 |
| Seagate | Momentus | 5400 | IDE 133 |
| Hitachi | 39T2525 | 7200 | ATA 133 |
| Hitachi | 38L2392 | 10000 | SCSI |

| Product Description | RPM | Int. Type |
|---|---|---|
| Seagate Barracuda HD | 5400 | ATA 100 mb/s |
| WD RaptorHDD | 7200 | IDE 133 mb/s |
| Seagate Momentus | 5400 | IDE 133 mb/s |
| Hitachi model 39T2525 | 7200 | ATA 133 mb/s |

**(b) Similar attributes have similar value distributions**

| Brand | Model | Speed | Interface |
|---|---|---|---|
| Seagate | Barracuda | 5400 | ATA 100 |
| Western Digital | Raptor | 7200 | IDE 133 |
| Seagate | Momentus | 5400 | IDE 133 |
| Hitachi | 39T2525 | 7200 | ATA 133 |

| Product Description | RPM | Int. Type |
|---|---|---|
| Seagate Barracuda HD | 5400 | ATA 100 mb/s |
| WD RaptorHDD | 7200 | IDE 133 mb/s |
| Seagate Momentus | 5400 | IDE 133 mb/s |
| Hitachi model 39T2525 | 7200 | ATA 133 mb/s |

**(c) Determining attribute correspondences utilizing values**

| Speed | 5400,7200,5400,7200 |
|---|---|
| Interface | ATA, 100, IDE, 133, IDE, 133, ATA, 133 |

| RPM | 5400 ,7200,5400,7200 |
|---|---|
| Int. Type | ATA, 100, mb/s, IDE, 133, mb/s, IDE, 133, mb/s, ATA, 133, mb/s |

**(d) Divergence scores**

| Catalog Attribute | Merchant Attribute | Jensen-Shannon Divergence |
|---|---|---|
| Speed | RPM | 0.00 |
| Speed | Int. Type | 0.69 |
| Interface | RPM | 0.69 |
| Interface | Int. Type | 0.13 |

terms based on the assumption that similar terms have similar "contexts" (*i.e.,* words appearing around the term in a corpus). Lee [12] compared multiple measures of distributional similarity and concluded that, for the synonym detection problem, the best performing ones are Jensen-Shannon (JS) divergence [13] and Jaccard coefficient [9]. Given two attributes $A$ and $B$, the JS divergence is computed as $JS(p_A \| p_B) = \frac{1}{2}KL(p_A \| p_M) + \frac{1}{2}KL(p_B \| p_M)$, where $p_M$ is the "average" distribution between $p_A$ and $p_B$, computed as $p_M = \frac{1}{2}p_A + \frac{1}{2}p_B$; and $KL(p_A \| p_M)$ corresponds to the Kullback-Leibler (KL) divergence [11] between $p_A$ and $p_M$:

$$KL(p_A \| p_M) = \sum_t p_A(t) log \frac{p_A(t)}{p_M(t)}$$

KL measures the difference between the probability distributions of $A$ and $M$. The Jaccard coefficient considers only counts for the different terms, and it is computed as $J(\mathcal{A}, \mathcal{B}) = \frac{|\mathcal{A} \cap \mathcal{B}|}{|\mathcal{A} \cup \mathcal{B}|}$.

Let us continue with the example of Figure 5. In Figure 5(b), we have a set of offers and their matching products. In Figure 5(c), we show the bag of words for the attributes Speed and Interface from the catalog, and RPM and Int. Type from the merchant. Suppose that we want to find an attribute correspondence for the catalog attribute Interface. Intuitively, the correct correspondence should include the merchant attribute Int. Type. In fact, as Figure 5(c) shows, although the word distributions of Interface and Int. Type are not exactly the same, they are more similar to each other than Interface and RPM. This is also reflected in the measures of distributional similarity. In Figure 5(d), we show the JS divergence for different combinations of attributes and we can see that the attributes Interface and Int. Type are closer to each other than the attributes Interface and RPM (for the former it is 0.13, and 0.69 for the latter).

We compute distributional similarity features for each tuple $\langle A^p, A^o, M, C \rangle$ where $C$ is a category in the catalog's taxonomy, $M$ is a merchant, $A^p$ is an attribute from the product specification for category $C$, and $A^o$ is an attribute from an offer specification from merchant $M$ and category $C$. For each feature, we first select a set of products $P$ and offers $O$ (the details of the product and offer selection depend on the feature and is described below). We then assemble a bag of words $\mathcal{A}^p$ that contains all the values for attribute $A^p$ of products of $P$; and a bag of words $\mathcal{A}^o$ that contains all the values for attribute $A^o$ of offers of $O$. We compute the distribution $p_{A^p}$ for each term $t$ as follows (analogously for $p_{A^o}$):

$$p_{A^p}(t) = \frac{\text{number of times } t \text{ appears in } \mathcal{A}^p}{\text{total number of elements in } \mathcal{A}^p}$$

| Name | Similarity measure | Grouping |
|---|---|---|
| `JS-MC` | JS | Merchant and Category |
| `JS-C` | JS | Category |
| `JS-M` | JS | Merchant |
| `Jaccard-MC` | Jaccard | Merchant and Category |
| `Jaccard-C` | Jaccard | Category |
| `Jaccard-M` | Jaccard | Merchant |

**Table 1: Features used by the classifier**

To capture the different associations (and potentially meaningful value distributions) between offers and products, we consider different ways to group the matching entities. Each grouping leads to a different feature for the classifier.

• *Group by merchant and offer:* This corresponds to the example above, where we matched all the offers of a given merchant in the Hard Drives category to the hard drive products in the catalog matching those offers. For a tuple $\langle A^p, A^o, M, C \rangle$, we compute the distributional similarity measures on the set of offers $O$ of merchant $M$ in category $C$; and the set of products in the catalog that match to offers of $O$. The features computed from these sets provide a powerful signal when there are enough offers for the merchant $M$ and category $C$. However, if the data is sparse, the signal may be weak. To tackle this problem, we also consider features computed on larger groups (group by merchant only, or group by category only), which we explain next.

• *Group by category:* The reason to consider this grouping is that, within a category, if a merchant gives an interpretation to an attribute, other merchants are likely to give the same interpretation. For example, if the attribute "MPN" means "Model Part Number" to one merchant, it is likely to mean the same to the other merchants. The features based on "group-by-category" are computed as follows. For a tuple $\langle A^p, A^o, M, C \rangle$, we compute the distributional similarity measures on the set of offers $O$ in category $C$; and the set of products in the catalog that match to offers in $O$. Notice that this grouping is coarser than the previous one, since the set $O$ contains offers for every merchant in category $C$.

• *Group by merchant:* A given merchant is likely to give similar interpretations to the attributes in different categories. For example, if the attribute "MPN" means "Model Part Number" for merchant Buy.com in category "Cameras", it is likely to mean the same to this merchant in any other category. We derive features based on "group-by-merchant" as follows. Given a tuple $\langle A^p, A^o, M, C \rangle$, we compute the distributional similarity measures on the set of offers $O$ for merchant $M$; and the set of products in the catalog that match to offers in $O$.

### 3.2 Automated Training Set Creation

We employ a classifier that uses logistic regression [8] to predict whether a candidate $\langle A, B, M, C \rangle$ tuple is actually an attribute correspondence. Table 1 describes the features used by the classifier, which are associated to the different distributional similarity measures and groupings presented above. Notice that we consider all the combinations of similarity measures (JS divergence and Jaccard) and groupings of offers.

Like any supervised learning technique, logistic regression requires a training set. Typically, training sets consist of data manually labeled by domain experts. However, this is not an option in a product synthesis framework since the techniques must be applicable to thousands of categories and merchants. Thus, in our approach we construct an *automatically labeled training set*. The construction is based on assumptions that are natural in the product synthesis context. In particular, we leverage *name identity candidate tuples*, *i.e.,* candidate correspondences $\langle A, A, M, C \rangle$, where the catalog and the merchant use exactly the same name for the attribute. Intuitively, within a large corpus, these exact matches are likely to occur and when they occur, there is a high probability that the tuple is actually a valid attribute correspondence.

Before giving the details of the training set construction, let us illustrate it with an example. Suppose that the catalog has an attribute called "Resolution" in the category "Digital Cameras". Suppose that for the (historical) offers that match to the camera products, a merchant also has an attribute called "Resolution". Then, it is likely that these two attributes have the same meaning. Furthermore, for any given offer of the merchant, since we know that the attribute name "Resolution" is being used, it is unlikely that any other name would be used to refer to the resolution of the camera. This observation leads to the following assumptions:

- If $\langle A, A, M, C \rangle$ is a name identity candidate tuple, then it is an attribute correspondence.
- If there are candidate tuples $\langle A, A, M, C \rangle$ and $\langle A, B, M, C \rangle$ where $A \neq B$, then $\langle A, B, M, C \rangle$ is not an attribute correspondence, since we assume a merchant $M$ will use exactly one name to refer to the catalog attribute $A$.

Using these assumptions, we build the training set for the classifier as follows. Let $M$ be a merchant and $C$ be a category. Each name identity candidate tuple $\langle A, A, M, C \rangle$ is considered a positive example (*i.e.,* $label(\langle A, A, M, C \rangle) = 1$). If there are candidate tuples $\langle A, A, M, C \rangle$ and $\langle A, B, M, C \rangle$ where $A \neq B$, then $label(\langle A, B, M, C \rangle) = 0$. Notice that the labels are defined only for the cases when a name identity exists. The training set constructed in this way turns out to be effective for learning a high accuracy classifier, as shown in our experiments of Section 5.

## 4. RUN-TIME PROCESSING PIPELINE

As illustrated in Figure 4, the pipeline consists of four components which are described below.

**Attribute Extraction from Web Pages.** We leverage information on merchants' landing pages to extract a set of attribute-value pairs for offers. We have implemented a simple extractor that parses the DOM tree of the Web page and returns all tables on the page. It also selects the attribute-value pairs from the tables, *i.e.,* rows with two columns, where we consider the first column to be the attribute name and the second column to be the attribute value. Although this extractor misses offers that are not formatted as tables (*e.g.,* a bulleted list of attributes), since there are often several offers associated with a given product, it still obtains broad coverage. For the offers that are formatted as tables, due to the wide variation in table structure, the extractor will invariably make mistakes. But as we show in the experiments of Section 5, a major strength of our approach is that the Schema Reconciliation component can filter out much of the noise that results from incorrect extraction. The intuition behind this is that the distribution of values for the catalog attributes will be very different from the distribution of incorrectly extracted values. Thus, no attribute name correspondences would be learned during the Offline Learning phase for the noisy pairs.

**Schema Reconciliation.** This component applies the attribute name correspondences obtained during the Offline Learning Phase to the offer attribute-value pairs extracted from the merchant Web pages. Let $o$ be an offer for category $C$ and merchant $M$, and $\langle A, v \rangle$ be one of the attribute-value pairs extracted from the merchant's Web page. If $\langle B, A, M, C \rangle$ is an attribute correspondence produced by the Attribute Correspondence Creation component during the Offline Learning phase, then the Schema Reconciliation component outputs a pair $\langle B, v \rangle$. Otherwise, the pair $\langle A, v \rangle$ is discarded.

**Clustering.** By leveraging the correspondences obtained from the Schema Reconciliation, the Clustering component first extracts the key attributes (Model Part Number or universal identifier UPC) for each offer. Then, offers that have the same key are clustered together, leading to clusters that have a one-to-one correspondence to a product instance. For example, the key attribute Model Part Number may be represented in an offer as "MPN" and in another offer as "Mfr. Part #". The Schema Reconciliation component maps both offer attributes into Model Part Number, thus letting us compare keys of offers and correctly group them. It is worth noting that we group offers based on key attributes to ensure that each resulting cluster corresponds to one product instance. However, other strategies to cluster the offers could be applied (e.g, considering other key attributes).

**Value Fusion.** The goal of this component is to combine the information from multiple offers in a cluster to create a (single) product specification. Given a set of attributes in the schema of a catalog category, Value Fusion selects from the cluster a *representative* value for each attribute. In our system, we implemented a simple extension of majority voting that generalizes to the case of textual values. For more details, see the Appendix (Section A).

## 5. EXPERIMENTAL EVALUATION

In this section, we present an experimental evaluation of our approach. First, we validate the effectiveness of the end-to-end system. Then, we drill down into the schema reconciliation component and discuss the different design choices that were made.

### 5.1 Effectiveness of Product Synthesis

**Experimental Setup.** We evaluated our approach using the catalog and offer feeds from Bing Shopping. The data set consists of 856,781 offers from 1,143 different merchants and covering 498 catalog categories, including: computing products (laptops, hard drives, etc.), cameras (digital cameras, lenses, etc.), home furnishings (bedspreads, home lighting, etc.), and kitchen and housewares (air conditioners, dishwashers, etc.). In the Offline Learning phase, we built the classifier by automatically creating a training set of 76,635 elements, 16,213 of which are positive. We then ran the classifier on 414,476 candidate tuples and predicted 82,889 correspondences to be valid. Such correspondences were used by the Schema Reconciliation component during the Run-Time Offer Processing phase of the pipeline.

**Evaluation Methodology.** Creating a ground truth set for product synthesis is challenging since the generated product specifications may not exist in the catalog. Thus, we had to perform a laborious task of labeling the output of product synthesis based on information from *product manufacturers*. For each product specification to be evaluated, we asked the labelers to identify its corresponding product by looking at attributes such as Brand, Model, Model Part Number, and Manufacturer, if they were available. Then, they used the attributes to browse the Web site of the manufacturer to find the page that contains the product specification. If such page could not be found, the entire specification (all attribute-value pairs) was considered to be invalid. Otherwise, each attribute-value pair of the specification generated by our algorithm was evaluated against the manufacturer's specification. Notice that in this process of creating the ground truth, we are not resorting to the merchant's site but rather the actual manufacturer's site. The main reason is to avoid bias: since the product specifications to be evaluated were constructed using merchant data, we must use a different source for the ground truth.

| Input Offers | 856,781 |
|---|---|
| Synthesized Products | 287,135 |
| Synthesized Product Attributes | 1,126,926 |
| Attribute Precision | 0.92 |
| Product Precision | 0.85 |

Table 2: Quality of synthesized product specifications

|  | Cameras | Computing | Furnishing | Kitchen |
|---|---|---|---|---|
| Avg Attrs / Product | 4.34 | 5.11 | 1.12 | 1.4 |
| Attribute precision | 0.91 | 0.91 | 0.99 | 0.97 |
| Product precision | 0.72 | 0.79 | 0.99 | 0.95 |

Table 3: Synthesis per top-level category

|  | Attribute recall | Attribute precision |
|---|---|---|
| Products with $\geq 10$ offers | 0.66 | 0.89 |
| Products with $< 10$ offers | 0.47 | 0.91 |

Table 4: Precision and recall for synthesized attributes

We define *attribute precision* as the ratio of attribute-value pairs labeled as correct over the total number of attribute-value pairs in the labeled sample. We also define a metric of precision for the entire product. Our metric for product precision is very strict in the sense that we consider a product to be *correct* only when all the attributes that we synthesized for it are correct. In particular, we define *product precision* as the ratio of correct products over the total number of products in the sample.

To evaluate attribute recall, we compute what fraction of the attributes mentioned on the merchant pages actually appear in the synthesized products. To do so, we sampled the synthesized products $P$, and for each sampled product $p \in P$, we gathered the set of offers $O$ that were used to synthesize $p$. We inspected each offer $o \in O$, manually extracted and integrated the attributes and values from different offers, and constructed a product $p_{gt}$. The set of manually created products $P_{gt}$ was then used as ground-truth. Let $X$ be the synthesized attributes in $P$ and $Y$ be the set of attributes from products in $P_{gt}$. We define *attribute recall* as $\frac{|X \cap Y|}{|Y|}$. Notice that the ground truth corresponds to all possible product attributes that can be synthesized using data extracted from the merchant pages. Of course, this may not include all conceivable attributes for a product, but it does include a significant subset: the set of attributes that merchants deem as useful for their customers.

**Results.** We ran the pipeline over the input offers and obtained 287,135 products which contained 1,126,926 attribute-value pairs. We sampled 1,447 attribute-value pairs, corresponding to 400 products (which gives a confidence level of 95% based on interval estimation [14]). The results are summarized in Table 2. We obtain an attribute precision of 0.92 and a product precision of 0.85. This precision is remarkably high, specially if we consider that all pairs were automatically extracted from merchant Web pages. This indicates that the schema reconciliation component is indeed effective at removing noise.

Table 3 shows a breakdown of the results for the top-level categories in the taxonomy, which reflect the aggregated results over the 498 categories we considered. Notice that the products in the Cameras and Computing top-level categories have a higher average number of synthesized attributes. This is due to the nature of these categories, whose products have rich product specifications on the merchant Web pages. In contrast, products in categories such as Home Furnishings and Kitchen & Housewares (*e.g.*, bedspreads, kitchen utensils) have fewer attributes specified on the landing pages. While attribute precision is relatively high across categories (0.91 to 0.99), the product precision is high for some categories (0.95 and 0.99 for Home Furnishings and Kitchen & Housewares, respectively), and lower for others (0.72 and 0.79 for Cameras and Computing, respectively). This is an artifact of our strict notion of product precision: the likelihood of having at least one incorrect attribute is higher for products with large number of attributes (*e.g.*, Computing) as opposed to categories with a small number of attributes (*e.g.*, Home Furnishings).

To evaluate recall, we examined products with different offer set sizes, *i.e.*, products associated with large and small sets of offers. In particular, we created two samples, each with 30 products: one for randomly selected products associated with more than 10 offers each; and the other, with less than 10 offers each. This resulted in a total of 498 offers and 2808 attribute value pairs that had to be manually inspected to create the ground truth products. As Table 4 shows, while our approach leads to high and similar precision values for both sets (0.89 and 0.91), the recall values show a larger difference (0.66 versus 0.47). The reason is that when the number of offers is larger, there tends to be more evidence (candidates) from the merchants for each catalog attribute to be synthesized. To see this, consider the pool of attribute-value pairs extracted from the offers for each product (i.e., the union of all attributes from all merchants, including synonymous attributes). For the set of products with more than 10 offers, there were an average of 84.6 attribute-value pairs per product; while for the other, this number drops to 9. Also, for larger offer sets, more attributes can be synthesized. For the products with more than 10 offers, the average number of synthesized catalog attributes was 13.3; for products with less than 10 offers, the number of synthesized attributes drops to 3.1.

### 5.2 Effectiveness of Schema Reconciliation

**Experimental Setup.** To assess the effectiveness of schema reconciliation, we evaluate the quality of the attribute correspondences generated during the Offline Learning phase. We compare our approach, described in Section 3, against alternative configurations. In each case, the input is a set of products and offers, and the output is a set of candidate attribute correspondences. We measure the quality of a configuration in terms of the precision and coverage of its output. For each configuration, we sample the output and manually label the correspondences as correct or incorrect. Note that we sample from candidates $\langle A, B, M, C \rangle$, where $A \neq B$, *i.e.,* we exclude from the evaluation set the name identity correspondences which are used to construct the classifier. All configurations that we consider give a score $\theta$ to each output correspondence. We consider precision and coverage at different values of $\theta$. Let $Z$ be the set of correspondences output by a configuration with a score greater than $\theta$. Let $X$ be a sample of those correspondences. Let $Y$ be the subset of $X$ that contains all the correspondences of $X$ labeled as correct. Then, we define *precision at* $\theta$ as $\frac{|X \cap Y|}{|X|}$. We also measure the *coverage at* $\theta$ as the size of $Z$, the total number of correspondences in the output whose score is greater than $\theta$. We use $\theta$ as a parametric knob, and report precision as a function of coverage. Notice that while obtaining absolute recall numbers is not feasible at the scale of the problem that we are addressing, the notion of coverage at $\theta$ that we use is in fact an effective way to compare the *relative recall* [15] of different schema matching techniques. In particular, if an algorithm leads to higher coverage than another at the same precision, then it has higher recall. For more details, see Appendix B.

**Validation of Design Choices.** In what follows, we evaluate two major design choices in our schema reconciliation approach: the use of a classifier to combine similarity features based on different levels of aggregation; and the use of historical offer-to-product matches in order to compute distributional similarity features. To evaluate the effectiveness of the classifier, we compare our approach, which combines similarity features by grouping products and offers are grouped in different ways, against baselines where a

414

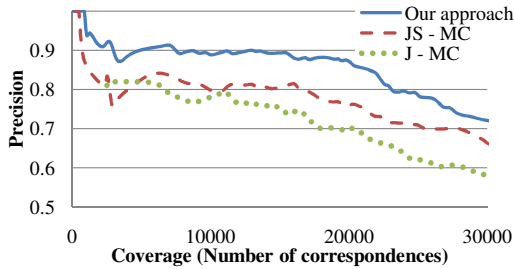

**Figure 6: Comparison of our schema reconciliation with baselines that do not combine distributional similarity.**

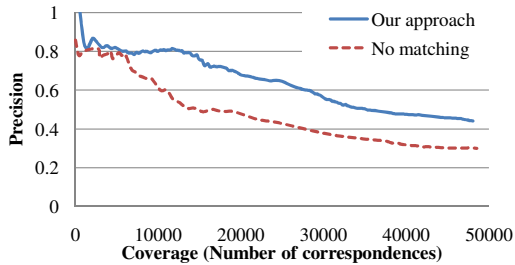

**Figure 7: Comparison of our schema reconciliation with baseline that does not employ historical instance matches.**

single similarity measure is used to score the candidate correspondences (thus no classifier is needed). In particular, we consider two baselines: one where only the feature `JS-MC` is used; and another where only the feature `Jaccard-MC` is used (see Table 1).

As in the previous section, the process for obtaining correspondences was applied to all categories and merchants. For each configuration, we sampled and labeled 384 correspondences from the output (which gives a confidence level of 95%). The results in Figure 6 show that our approach consistently outperforms the use of individual similarity measures. For example, we obtain 20K correspondences with 0.87 precision, while by using the features `JS-MC` and `Jaccard-MC` alone, precision drops to 0.76 and 0.69, respectively. This confirms the benefits of combining signals from multiples levels of aggregation using a classification-based approach.

To assess the benefit of historical offer-to-product matches, we compare our approach to a baseline where this information is not used. The baseline uses the same similarity measures (Jaccard and JS divergence) as our approach, but instead of considering only products that *match* to offers, it takes into account all products in a given category $C$ and all offers associated with $C$. We ran this experiment for 92 categories (corresponding to subcategories of Computing such as monitors, workstations, and mobile devices), covering 1,143 merchants. For each configuration, we sampled and labeled 384 candidate correspondences. As Figure 7 shows, our approach outperforms the configuration where historical offer-to-product matches are not used. This confirms our hypothesis that historical instance matches produce more accurate distributions, which, in turn, lead to higher-quality correspondences.

**Comparison against existing schema matching techniques.** We compared the results of our classifier against state-of-the-art schema matching techniques and systems: COMA++ [6], DUMAS [1], and the Naïve Bayes matcher used by LSD [5]. COMA++ is a publicly available schema matching framework which supports both name-based and instance-based matchers. DUMAS is a schema matching tool that leverages previously-known instance matches (*i.e.,* historical offer-to-product matches in our case) to improve the quality of schema matching. However, their technique is not classification-based and does not employ distributional similarity. Similar to our approach, the Naïve Bayes classifier of LSD also uses learning, but the features it uses are different from ours, in particular, it does not use distributional similarity. A detailed discussion of our implementation of DUMAS and the Naïve Bayes classifier of LSD is given in Appendix C.

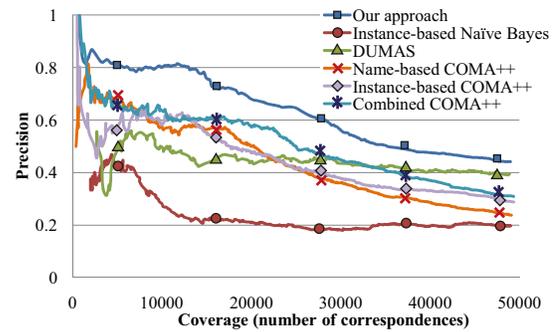

**Figure 8: Comparison of our schema reconciliation approach against other schema matching approaches.**

We generated correspondences for 92 categories (all computing categories) and 1,143 merchants. The results of the comparison are given in Figure 8. Our approach consistently outperforms all other configurations, and achieves significantly higher precision. For example, we obtain 10K correspondences with 0.8 precision, while the other approaches obtain precision values that vary between 0.28 and 0.6. Note that the higher coverage obtained by our approach implies that it has higher recall relative to the other configurations (for a detailed explanation, see Appendix B).

It is interesting to contrast the results for the different COMA++ configurations.[1] The instance-based COMA++ configuration obtains high precision for small coverage values, but then the precision drops significantly. In contrast, the name-based configuration has lower precision at low coverage, which can be explained by the fact that attributes with similar names are not necessarily matching attributes (*e.g.,* attributes "Memory Technology" and "Graphic Technology"). The combination of instance-based and name-based matching outperforms the other two configurations. It is worthy of note that while our schema reconciliation approach only takes instances into account, it outperforms the combined matchers for COMA++. It is likely that our approach would perform even better if we combined name and instance matches, which we leave as future work.

These results should not be construed as shortcomings of existing approaches, but rather as evidence that they were not designed with the requirements of product synthesis in mind. For example, LSD and COMA++ are general frameworks where different classifiers can be plugged in. And thus, if classifiers were created with category-specific features based on domain knowledge, they would likely lead to high precision and recall.

## 6. RELATED WORK

While there is an extensive literature on schema matching (for a survey, see [16]), schema matching is typically presented as a stand-alone component and not evaluated in the context of a complex framework such as product synthesis. One possible exception is the work by Wick et al. [19], which presented an approach where a single Conditional Random Field is built to handle schema matching, entity resolution, and data fusion. In contrast to our approach, Wick et al. require large amounts of training data and it has only been tested for a small number of (relatively small) schemas.

LSD [5] and COMA++ [6] are general frameworks for schema matching that allow the combination of multiple matchers. LSD

---
[1] We have also explored additional configurations using different parameter settings. We report these results in Appendix D.

uses learning classifiers (base learners), which can be trained to match attribute names and values, and a meta-learner to combine the predictions of the base learners. For our scenario, LSD would require training data in the form of manually-created correspondences between the the catalog and merchant schemas; and this is not feasible due to the large number of categories and merchants. COMA++ provides a library of matchers that use linguistic and terminological similarity between attribute pairs to identify matches (*e.g.,* edit distance, trigrams); and the ability to combine them. As our experiments show, for the product synthesis scenario, our schema reconciliation approach outperforms the standard name and instance similarity matchers used by COMA++.

Bilke and Naumann [1] proposed the DUMAS schema matching algorithm, which leverages the availability of duplicate records to help in the discovery of attribute correspondences. Their algorithm relies on the existence of elements that have very similar representations in their attribute-value pairs. In their paper, they presented experiments for apartment classified ads, and observed that, in the same time period, a given person may place ads for the same apartment on multiple newspapers. In general, this does not hold in product synthesis, where the representation of offers may differ even if they correspond to the same product, and they often also differ from the product schema in the catalog. The results reported in Section 5 provide evidence that redundancy alone is not sufficient to obtain high precision and recall for product synthesis. Dhamakar et al. [3] also make use of duplicate records, but they do so to refine initially discovered complex matches. Besides, unlike our approach, they do not leverage the overlap in data as part of the construction of value distributions.

Kang and Naughton [10] presented a schema matching technique that uses distributional similarity (mutual information) and graph matching. Their approach leverages correlations of attributes *within* a table. As such, it complements other approaches to schema matching that rely on attribute names and values. Investigating whether there are correlations of attributes within product or offer specifications is a direction that we intend to pursue for future work. Carlson et al. [2] proposed an approach to extract structured records from Web pages. Their approach is based on a measure of distributional similarity [12] and uses a learning classifier. There are important differences between their approach and ours: their features are designed to compare Web pages—in their work, there is no notion of a master, structured database (our catalog); and they do not exploit instance matchings to create the features.

In addition to matching database fields, schema matching techniques have been used for matching Web form elements (*e.g.,* [7, 17]). Although both form-schema matching and product synthesis have the requirement of matching a large number of instances, there are important differences between them. Web forms have schemata that are simpler than the ones in a product catalog, and while some form elements may have values associated with them (*e.g.,* selection lists), many have none (*e.g.,* text boxes). In contrast, attribute values play an important role in product synthesis. Although the form-matching approaches have addressed the problem of matching hundreds of forms, they report experiments for only a small number of domains and consequently, schemata. Finally, a limitation of the form matching approaches stems from their assumption that form schemata are clean and normalized: when data is noisy and many variations of a given attribute name are used, their effectiveness decreases [7]. For product synthesis, it is necessary to handle thousands of categories and merchants (each pair of category and merchant corresponding to a different schema), where there is a wide variation in both attribute names and values within a category.

# 7. CONCLUSIONS

We presented a scalable, end-to-end solution to the product synthesis problem. We used large-scale data from Bing Shopping to perform a thorough experimental evaluation, where we showed that our approach outperforms existing state-of-the-art techniques. There are several avenues we plan to pursue in future work. Because the attributes we synthesize come from merchant offers, which are often terse, we plan to investigate the use of Web pages from manufacturers as a means to enrich the attribute set. While offer pages can be easily retrieved from URLs provided in feeds, locating the manufacturers' sites and the relevant product pages within these sites poses additional challenges. We would also like to integrate other matchers with our framework, notably, name matchers.

**Acknowledgments.** We would like to thank Daniel Keller, Arun Sacheti, Radek Szamrej, David Talby, and Rama Shenai for insightful discussions and feedback; Anitha Kannan and Fabian Suchanek for their comments on different aspects of this work; Huong Nguyen for her help in creating our gold data sets; and Erhard Rahm and Sabine Maßmann, for sharing with us the latest version of COMA++ and for their help in configuring the system for our experiments.

# 8. REFERENCES

[1] A. Bilke and F. Naumann. Schema matching using duplicates. In *ICDE*, pages 69–80, 2005.
[2] A. Carlson and C. Schafer. Bootstrapping information extraction from semi-structured web pages. In *ECML PKDD*, pages 195–210, 2008.
[3] R. Dhamankar, Y. Lee, A. Doan, A. Halevy, and P. Domingos. Imap: discovering complex semantic matches between database schemas. In *ACM SIGMOD*, pages 383–394, 2004.
[4] H.-H. Do and E. Rahm. Coma - a system for flexible combination of schema matching approaches. In *VLDB*, pages 610–621, 2002.
[5] A. Doan, P. Domingos, and A. Halevy. Reconciling schemas of disparate data source: a machine learning approach. In *SIGMOD*, pages 509–520, 2001.
[6] D. Engmann and S. Maßmann. Instance matching with coma++. In *BTW Workshops*, pages 28–37, 2007.
[7] B. He and K. C.-C. Chang. Automatic complex schema matching across web query interfaces: A correlation mining approach. *TODS*, 31(1):346–395, 2006.
[8] D. Hosmer and S. Lemeshow. *Applied logistic regression (Wiley Series in probability and statistics)*. Wiley-Interscience Publication, 2000.
[9] P. Jaccard. The distribution of the flora in the alpine zone. *New Phytologist*, 11(2):37–50, 1912.
[10] J. Kang and J. Naughton. On schema matching with opaque column names and data values. In *SIGMOD*, pages 205–216, 2003.
[11] S. Kullback. *Information Theory and Statistics*. Dover, 1968.
[12] L. Lee. Measures of distributional similarity. In *COLING*, pages 25–32, 99.
[13] C. Manning and H. Schütze. *Foundations of Statistical Natural Language Processing*. The MIT Press, 1999.
[14] W. Mendenhall. *Introduction to probability and statistics*. Duxbury Press, 1986.
[15] P. Pantel, D. Ravichandran, and E. Hovy. Towards terascale knowledge acquisition. In *COLING*, pages 771–777, 2004.
[16] E. Rahm and P. Bernstein. A survey of approaches to automatic schema matching. *VLDBJ*, 10(4):334–350, 2001.
[17] W. Su, J. Wang, and F. Lochovsky. Holistic query interface matching using parallel schema matching. In *EDBT*, pages 77–94, 2006.
[18] J. Weeds, D. Weir, and D. McCarthy. Characterising measures of lexical distributional similarity. In *COLING*, page 1015, 2004.
[19] M. Wick, K. Rohanimanesh, K. Schultz, and A. McCallum. A unified approach for schema matching, coreference and canonicalization. In *KDD*, pages 722–730, 2008.

# APPENDIX
## A. DETAILS OF VALUE FUSION

The goal of this component is to combine the information from multiple offers in a cluster to create a (single) product specification. Given a set of attributes in the schema of a catalog category, Value Fusion selects from the cluster a *representative* value for each attribute. One approach for choosing the representative value is majority voting. For example, suppose that we have a cluster with five offers, four of which contain the attribute-value pairs ⟨ Memory Capacity, 1024 ⟩, and one of which contains the attribute value pair ⟨ Memory Capacity, 2048 ⟩. Then, a majority voting approach would choose the value 1024.

While majority voting works well for values that consist of exactly one token, it may not be appropriate for textual values that involve multiple tokens. For example, suppose that we have offers containing the attribute-value pairs ⟨ Operating System, Windows Vista ⟩, ⟨ Operating System, Microsoft Windows Vista ⟩ and ⟨ Operating System, Microsoft Vista ⟩. The three values are different, and thus majority voting does not favor any of them. However, intuitively, we would like to choose the value "Microsoft Windows Vista" because it contains the word "Vista", which appears in the three offers, and the words "Microsoft" and "Windows", which appear in two of the offers. We capture this intuition with a generalization of majority voting at the term level.

The generalization of majority voting is done as follows. Consider a cluster of offers that contains attribute-value pairs. Suppose the attribute-value pairs of an attribute A is $\langle A, v_1 \rangle, \ldots, \langle A, v_n \rangle$. Let $T$ be the set of terms that appear in $v_1, \ldots, v_n$. We construct a vector $X$ whose dimension is $|T|$ and where each position is associated to one of the terms of $T$. We then compute the centroid for the set of vectors and choose as representative value the one that is closest (in terms of Euclidean distance) to the centroid. Continuing our example, let $v_1$ be "Windows Vista", $v_2$ be "Microsoft Windows Vista", and $v_3$ be "Microsoft Vista". We then create 3-dimensional vectors for each value, where the first term corresponds to the word "Microsoft", the second term corresponds to the word "Windows" and the third term corresponds to the term "Vista". The vectors are $\langle 0, 1, 1 \rangle$ for $v_1$; $\langle 1, 1, 1 \rangle$ for $v_2$; and $\langle 1, 0, 1 \rangle$ for $v_3$. The centroid $C$ is $\langle \frac{2}{3}, \frac{2}{3}, 1 \rangle$. The distance to the centroid for $v_1$, $v_2$ and $v_3$ is 0.75, 0.47 and 0.75, respectively. We conclude that the closest value is $v_2$ and we choose it as the representative value.

## B. COVERAGE AND RELATIVE RECALL

While it is not feasible to compute the absolute recall for the attribute correspondences, it is possible to measure *relative recall*, *i.e.,* the recall of an algorithm relative to another [15]. Let $R_A$ and $R_B$ be the recall of algorithms $A$ and $B$, respectively. The relative recall of algorithm $A$ to algorithm $B$ is defined as the ratio of $R_A/R_B$.

We now show that at the same level of precision $p$, if $A$ has higher coverage than $B$, then $R_A > R_B$. Suppose that we are interested in the recall of $A$ and $B$ at precision $p$. Let $\theta_A$ and $\theta_B$ be the scores at which algorithms $A$ and $B$ achieve precision $p$. Then, if the coverage of $A$ at $\theta_A$ is larger than the coverage of $B$ at $\theta_B$, $A$ has a higher recall relative to $B$ at precision $p$. To see why, let $c_A$ and $c_B$ be the coverage of $A$ and $B$ at precision p, respectively. Then, we can estimate the number of correct correspondences retrieved by $A$ as $c_A \times p$, and the number of correct correspondences retrieved by $B$ as $c_B \times p$. Let $y$ be the total number of correct correspondences in the ground truth. Then, the recall of $A$ is $\frac{c_A \times p}{y}$, and the recall of $B$ is $\frac{c_B \times p}{y}$. It follows that, if $c_A > c_B$, then the recall of $A$ is larger than the recall of $B$.

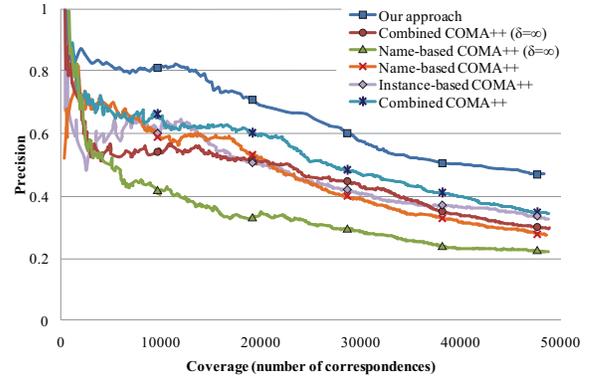

**Figure 9: Comparison of our schema reconciliation with COMA++ using $\delta = 0.01$ (default configuration) and $\delta = \infty$.**

## C. IMPLEMENTATION OF DUMAS AND NAÏVE BAYES MATCHERS

Since both our approach and DUMAS leverage previously-known instance matches, we set up the experiments so that both of them have access to the same offer-to-product matches. We implemented DUMAS by following the description provided in [1]. For each category $C$, with each product $p = \{a_1, \ldots, a_n\}$ in $C$ and an offer $o = \{b_1, \ldots, b_m\}$ of a merchant $M$ that belongs to $p$, we first computed a $m \times n$ similarity matrix $S_k$ that stores the similarity of each pair of field values $a_i$ and $b_j$ using the SoftTFIDF measure. A matrix $S_k$ is generated for each association of a product with an offer of $M$. The overall similarity matrix $S_M$ of merchant $M$ is produced as the average of all $S_k$:

$$S_M = \frac{1}{T} \sum_{k=1}^{T} S_k$$

where $T$ is number of associations of a product with an offer of $M$. We then use $S_M$ as input for the bipartite weighted matching problem and solve it to find the maximal matching $Match = \{\langle A, B, M, C \rangle\}$. Each match $\langle A, B, M, C \rangle$ in the maximal matching is treated as a candidate correspondence, where $A$ is a catalog attribute and $B$ is an attribute of merchant $M$.

For our experiments, we have also implemented the Naïve Bayes classifier used as an instance-based matcher in LSD [5]. We constructed a multi-class Naïve Bayes classifier for each category in the catalog (categories in our setting are analogous to tables in the integration setting of LSD). In particular, for a category, the classifier uses attribute names as class labels (*i.e.,* each attribute of the category is a class). The feature set of the classifier consists of every term that appears in any product instance in the category. The classifier is trained using the entire content of the catalog and is executed over all offers. For each category $C$, let $w_i$ be a term and $c_j$ be a class label. In the training phase, we estimate $P(c_j|w_i)$ as $n(w_i, c_j)/n(c_j)$, where $n(w_i, c_j)$ is the number of times term $w_i$ appears in attribute $c_j$ and $n(c_j)$ is the number of product instances in the catalog that have attribute $c_j$. Let $\langle B, v \rangle$ be some attribute-value pair for some offer of merchant $M$ in $C$. In the run-time phase, we compute the probability that $v$ belongs to class $A$ as $P(A|v) = P(v|A)P(A)$. Treating $v$ as a bag of words $\{w_1, \ldots, w_n\}$ and assuming that $w_i$ appears in $v$ independently of each other given $A$, then we have $P(v|A) = P(w_1|A) \ldots P(w_n|A)$. Now let $V$ be the set of all values that appear in some attribute-value pair $\langle B, v \rangle$ of merchant $M$ in category $C$ over all input offers, the score of $\langle A, B, M, C \rangle$ is computed as:

$$score(\langle A, B, M, C \rangle) = \frac{\Sigma_{v \in V} P(A|v)}{|V|}$$

Then, for each $A$, $M$, and $C$, a correspondence $\langle A, B, M, C \rangle$ is created if $score(\langle A, B, M, C \rangle) > score(\langle A, B', M, C \rangle)$ for every attribute $B'$ of the schema of merchant $M$ in category $C$.

## D. ADDITIONAL COMA++ EVALUATION

To further compare the effectiveness of our Schema Reconciliation component against COMA++, we evaluate COMA++ configurations with different values of $\delta$ [4]. In this experiment, besides using the default value $\delta = 0.01$, we also set $\delta = \infty$, which results in obtaining every possible pair of attributes as candidates, and then rank them by the score. As shown in Figure 9, our approach always lead to higher precision at the same level of coverage than all configurations of COMA++; and the results obtained by COMA++ using the default value of $\delta$ have higher precision than when using parameter $\delta = \infty$. This could be explained as with $\delta = 0.01$, we limit COMA++ to pick fewer correspondences per attribute, which improves precision at the cost of a reduced relative recall.